# IMPLEMENTING SCRUM TO DEVELOP A CONNECTED ROBOT

**Diego Armando Díaz Vargas[1], Rui Xue[2], Claude Baron[1], Philippe Esteban[1],**
**Rob Vingerhoeds[3], Citlalih Y. A. Gutierrez Estrada[4], Chao Liu[2]**

[1]LAAS-CNRS, Université de Toulouse, CNRS, INSA, UPS, 7 av. du Colonel Roche, 31400 Toulouse, France
{dadiazva, claude.baron, philippe.esteban}@laas.fr

[2]School of Economics and Management, Beijing University of Technology, 100 Ping Le Yuan, Chaoyang District, Beijing 100124, China,
{xuerui, liuchao}@bjut.edu.cn

[3]Institut Supérieure de l'Aéronautique et de l'Espace, 10 av. Edouard Belin BP 54032, 31055 Toulouse, France
rob.vingerhoeds@isae-supaero.fr

[4]Instituto Tecnológico de Toluca, Toluca, México,
citlaligh@yahoo.com

**ABSTRACT:** *Agile methods are receiving a growing interest from industry and these approaches are nowadays well accepted and deployed in software engineering. However, some issues remain to introduce agility in systems engineering. The objective of this paper is to show an agile management implementation in an educational project consisting in developing a connected mobile robot, and to evaluate the issues and benefits of adopting an agile approach. Among the most famous agile management methods, SCRUM has been chosen to lead this experiment. This paper first presents the project and how students traditionally manage it, then it describes how Scrum could be used instead. It evaluates the difficulties and interests to introduce agility in this project, and concludes on the ability of Scrum to design, test and progressively integrate the system, thus providing an operational prototype more quickly.*

## 1 INTRODUCTION

Every year, the Institut National des Sciences Appliquées de Toulouse (INSA Toulouse, Université de Toulouse, France) proposes a project to master students with an objective to design and realize a remotely controlled LEGO Mindstorms robot. The six to eight students project team is supervised by a professor, in charge of explaining the robot mission and helping with the management of the three-months project. The students' assignments are to define the necessary resources and contributing systems, and to organize their work and their team in order to be able to deliver the robot on time.

The project includes the development of the robot as well as the writing of a complete report, in which technical and organizational aspects have to be detailed. Students traditionally follow the V-cycle model to develop the system. The project is divided into six stages: initiation, preliminary development, principal development, robot realization, test and close. Although the V-cycle is a widely used reference in the industry, with the emergence of agile approaches, many project managers have paid attention to agile management of projects, essentially in software engineering projects. Therefore, students must be trained to agile management before joining the industry.

This article aims at deploying, testing and evaluating the issues and benefits of an agile approach on this project, in order to determine if it could be a candidate to train students to a basic practice of agile management. To this purpose, this paper considers and compares two options to implement agility.

The paper is organized as follows. Section 2 presents the way students are currently managing their project. Section 3 outlines the principles of the agile approach. Section 4 explains how the students could handle the project in an agile way. Section 5 concludes on the feasibility and interest of training students to an agile management on this project.

## 2 DEVELOPMENT OF A CONNECTED MOBILE ROBOT

The objective of the project is to deliver an autonomous remotely supervised wheeled robot. It includes the design and construction of the robot using LEGO building blocks. The robot can either be remotely operated or move autonomously on a delimited area. This area contains two distinct zones: a 'rest zone' where the robot is initially located, a 'camp' where the robot has several tasks to accomplish.

The mission of the robot is decomposed into three phases. It first has to reach the camp by autonomously following a line traced on the ground and to signal the operator that it arrived (send a 'bip'). There, an operator remotely takes the control of the robot to realize the tasks assigned to the robot. These tasks consist in dropping several wooden tokens that are embedded in a storage module in the robot, to different predefined zones. During the remotely controlled part of the mission the operator can visualize the robot environment thanks to a camera in order to avoid the different traps on the area (crevasses, low light areas, low Wi-Fi/Bluetooth areas) to drive the robot following the best route. Once the job accomplished, the robot returns to the rest zone.

From a given set of robot requirements, students have to define a logical and physical architecture for the robot, to construct it and to test it against its specifications. They first have to organize, divide and plan the project before dealing with more technical aspects (system analysis, conception, integration, tests, etc.). The project team has three months to lead the project.

The project team defines six stages to structure the project, with 7 milestones (named M1 to 7 in Figure 1).

| Initiation | Preliminary development | Principal Development | Robot realization | Test | Close |
|---|---|---|---|---|---|
| - Develop project charter | - System analysis<br>- Architectural conception of system | - Detailed requirement definition<br>- Architectural conception of system<br>- Reference configuration<br>- Integration<br>- Validation<br>- Verification | - Development of prototype<br>- Improvement system performance | - Test<br>- Maintenance | - Close project |
| M1 | M2 | M3 | M4 | M5 | M6   M7 |

Figure 1: The six stages of the project

**Initiation:** organization of the project group and definition of the responsibilities of each team member.
**Preliminary development:** definition of requirements, of alternatives for the robot architecture and of test cases.
**Principal development:** definition of the specifications and of the architectures of sub-systems; integration, validation and verification.
**Robot realization:** development of a robot prototype with LEGO bricks, improvement of the robot performance.
**Test:** students define test cases to verify the robot behavior and performance.
**Close:** students have to close the project before the deadline.

To lead the project, students use traditional methods, based on a predefined schedule, with cost constraints and quality objectives. They completely plan everything from the beginning, however wisely including some margins. The robot requirements are initially given to the students, as well as some physical constraints, such as the size and shape of tokens for instance. During the development stages, changes are not expected, or even allowed. This kind of management method is adapted to a V-model development cycle.

## 3   AGILE MANAGEMENT OF PROJECTS

Compared with the traditional management method used by the students, the agile management method attracts a lot of attention in industry because of several assets: flexibility, transparency, increased customer satisfaction, close communication between developers, etc. This section first compares traditional and agile management, then outlines the main advantages of agile management, to end with a brief introduction to the three mostly used agile methods.

### 3.1   Traditional management vs Agile management

The traditional management was characterized by Chin (2004), Fowler (2000) and Highsmith (2009) as a methodology whose emphasis is on detailed planning and resistance to change. Schwaber (2007) corroborates that about 50 percent of the time is spent on requirements, architecture and specification, and that all this is done before even building any functionality when the project management team used traditional practices. Agile management is described as an approach that aims for flexibility, simplicity, iterating in short periods of time, and incrementally adding value (Boehm, 2002), (Chin, 2004), (Cockburn, 2002), (Cohn and Ford, 2003), (Highsmith, 200), (Misra et al., 2010). A comparison conducted by Shenhar and Dvir (2007) highlights the main differences between the traditional and agile management method. Hoda et al. (2010) also compare them with a software development focus, and highlight the main differences.

### 3.2   Advantages of agile management

Agile management now is widely spread and especially applied in software development. Its use allows companies to quickly react to project changes and to adjust their processes and organizations. The use of agile management can also help companies to improve their success rate (Augustine et al., 2005) (Serrador and Pinto, 2015). The case study that has been carried out in the research of Serrador and Pinto (2015) has demonstrated that the use of agile management method can help the companies to end projects on time, on budget, with a high customer satisfaction. Furthermore, agile project management can solve the problems induced by a change of requirements (Azanha et al., 2017). Azanha et al. also states that agile management practices can help companies in getting value for them and their customers by the benefits obtained with the agile approach usage, applied to systems development. Nowadays, project leaders and teams find themselves in an environment disrupted by exponential advances in technology and demands from customers for more immediate delivery of value, agile techniques and approaches effectively manage disruptive technologies (PMI, 2017). Comparisons between agile management and traditional management in literature mainly focus on the software development context. This paper analyzes the advantages of agile management in the context of a mobile connected robot, where software is embedded with physical components.

### 3.3 Introduction to the three main agile management methods

Different agile methodologies are currently practiced in the industry. This subsection briefly introduces the most widely used. Among them SCRUM has been chosen for this project; it will be presented with more details.

*XP*
XP is an agile software development methodology created by Kent Beck and first presented in 1999, in a book titled Extreme Programming Explained: Embrace Change. According to the author, "the goal of eXtreme Programming is outstanding software development at lower cost, with fewer defects, higher productivity, and much higher return on investment" (Beck, 2004). XP does not rely on planning, analyzing and designing for the far future. Instead, it allows these activities to be made throughout the whole development, in short cycles (Beck, 1999).

*UPEDU*
UPEDU is a software development process specialized for education, developed by Pierre-N. Robillard, Patrick d'Astous of the École Polytechnique de Montréal, and Philippe Kruchten from Rational Software. UPEDU is an academic customization of the Rational Unified Process (RUP). RUP is an iterative software development process framework created by the Rational Software Corporation, an IBM division since 2003 (Robillard et al., 2001). Many companies claimed that they have applied agile management on their projects. However, Schwaber et Sutherland suggests only 30% of scrum-adopting companies will become "excellent development organizations". Change experts like Harvard Professor John Kotter (Kotter, 2009) regularly say 70% of major change efforts fail. Some of agile management method implementation in the companies only remain some aspects of agile management, such as its iteration or collaboration.

*Scrum*
Scrum was first proposed by Jeff Sutherland and Ken Schwaber in 1993 (Cervone, 2011), (Misra et al., 2010). It was originally designed for software development. Now the application of this method is not restricted to software development; it also can be applied to a large variety of domains, such as manufacturing, marketing, advertising agencies, architectural projects and banks (Cardozo et al., 2010). Among the three main agile management methods, Scrum is widely deployed in industry. In addition, Scrum is quite simple to train students without asking them to develop an advanced expert level beforehand. We thus chose Scrum to evaluate the interests of implementing agility into our educational project.

Scrum is founded on empirical process control theory, or empiricism. Empiricism asserts that knowledge comes from experience and making decisions based on what is known (Schwaber et al., 2017). Scrum is therefore defined as a framework which people can address complex adaptive problems, while productively and creatively delivering products of the highest possible value. This framework consists of scrum teams and their associated roles, events, artifacts and values, each component serves a specific purpose and is essential to Scrum's success and usage. Below a brief description of the different Scrum aspects is given (adapted from Schwaber et al., 2017):

Scrum Values

The Scrum values are followed by everyone in the project. It enumerates five values:

1) $V_1$ - *Courage*, Scrum Team members have courage to do the right thing and work on tough problems,
2) $V_2$ - *Focus*, everyone focuses on the work of the Sprint and the goals of the scrum team,
3) $V_3$ - *Commitment*, people personally commit to achieving the goals of the Scrum Team,
4) $V_4$ - *Respect*, Scrum Team members respect each other to be capable, independent people, and
5) $V_5$ - *Openness,* the Scrum Team and its stakeholders agree to be open about all the work and the challenges with performing the work.

Scrum Events

Scrum Events are specifically designed to enable critical transparency and inspection. Each event is a formal opportunity to inspect and adapt something. All events are time-boxed events and they may ensure an appropriate amount of time without allowing waste in the process. Scrum enumerates five events:

1) Sprint – a time-box of one month or less, during which a usable and potentially releasable product increment is created; an increment is the sum of all the items completed during a sprint and the value of the increments of all previous Sprints. The Sprint is the heart of Scrum and a new Sprint starts immediately after the conclusion of the previous Sprint.
2) Sprint Planning – a time-box to a maximum of eight hours for a one-month Sprint. Sprint planning plans the work to be perform in each Sprint. This plan is created and maintained by the collaborative work of all members of the project. Sprint planning also defines the sprint goal, an objective set for the Sprint, that can be met through the implementation of the Product Backlog.
3) Daily Scrum – a 15-minute time-boxed event, held every day of the Sprint. Daily Scrum is use to inspect progress toward the Sprint Goal and to inspect how progress is trending toward completing the work in the Sprint Backlog.
4) Sprint Review – the inspection of the increment. During this inspection stakeholders review the

results of a Sprint. A Sprint Review is held at the end of the Sprint and adapts the Product Backlog if needed.
5) Sprint Retrospective – an opportunity for the team to review itself and create a plan for improvements to be done during the next sprint. Sprint Retrospective occurs after the Sprint Review and prior the next Sprint Planning.

The Scrum Team

Scrum Teams have as a characteristic "self-organizing and cross-functional". Self-organizing teams choose how best to accomplish their work, rather than being directed by others outside the team. Cross-functional teams have all competencies needed to accomplish the work without depending on others. Furthermore, the Scrum Team delivers products iteratively and incrementally, maximizing opportunities for feedback. A Scrum Team comprises three entities:

1) a Product Owner: one person only and its decisions have to lead the entire team and the organization,
2) a Development Team: professionals who do the work of delivering a potentially releasable increment of product at the end of each sprint, and
3) a Scrum Master: it is responsible for promoting and supporting scrum framework, he does this by helping everyone understand practices, rules, values and scrum theory.

Scrum Artifacts

Scrum Artifacts represent work or value to provide transparency and opportunities for inspection and adaptation; they are designed to maximize transparency of key information. Product Backlog is considered as an ordered list of everything that is known to be needed in the product; the Product Owner is responsible for the Product Backlog. Sprint Backlog is the set of Product Backlog items selected for the Sprint, a plan for delivering the product increment and realizing the Sprint Goal; it makes the work that the Development Team identifies as necessary to meet the sprint goal visible. The Increment indicates the sum of all the Product Backlog items finalized during a Sprint, as well as the ones finished during all previous sprints. Product Backlog, Sprint Backlog and Increment are considered as artifacts.

Figure 2 shows the complete Scrum Framework and how each entity is distributed. The framework begins with the Product Backlog; the Product Owner defines all the features that the Development Team has to do and then to prioritize. The next step is the Scrum Planning; here the Product Owner and the Development Team plan which features of the Product Backlog have to be done and the time of each Sprint, which results in the Sprint Backlog. The Development Team has to do each Sprint defined in the Sprint Backlog. They have to hold daily meetings to organize the work that has to be done. Sprint Review allows the Development Team to solve the problems that occurred during the sprint development.

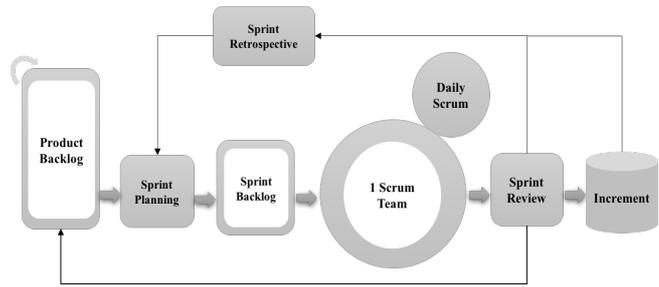

Figure 2: Global Scrum Framework (scrum.org, 2017)

## 4 USING AGILITY IN THE ROBOT PROJECT

Our proposal is to introduce agility in three project phases: principal development, robot realization and test phases (see Figure 1). This section considers and discusses several options to implement Scrum in these phases.

### 4.1 Implementing Scrum in the project

The robot mission is to move in a predefined environment, autonomously or remotely operated, as mentioned in Section 2. More specifically the robot shall achieve its mission following three successive phases:

- Phase 1: The robot shall move autonomously from the 'rest zone' to the 'camp' and send an audible signal once it arrives at the 'camp'.
- Phase 2: The robot shall drop tokens in predefined zones under the operator directions, avoiding traps on the route (one crevasse, two low light areas and three low Wi-Fi/Bluetooth areas).
- Phase 3: The robot shall join the 'rest zone' under the operator commands from an operating terminal.

These three phases can be considered as a Product Backlog and can be used to initiate the Sprint Planning. Before starting the Sprint Planning, students should decide who is going to be the Product Owner, the Scrum Master and who will contribute to the Development Team; the goal of this role definition is to focus on the scrum framework recommendations. Students can define the global architecture of the robot using the given requirements, that allows them to identify the different interactions between stakeholders and the robot. The robot can be decomposed into three subsystems as shown in Figure 3: the robot structure and motion subsystem, the vision subsystem and the Human Machine Interface subsystem.

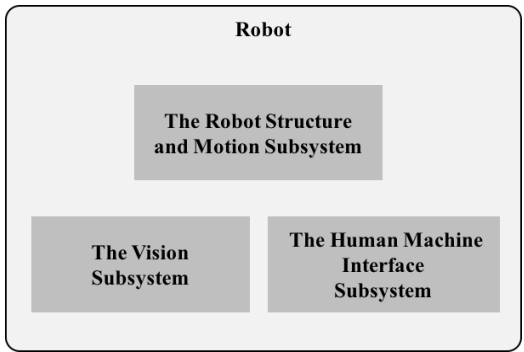

Figure 3: Global architecture of the robot

The robot structure and motion subsystem corresponds to the physical architecture of the robot with the motion subsystem; it has to answer all requirements of the specifications. The vision subsystem will allow an optimal view to the robot to accomplish its mission and the human machine interface subsystem will allow to remotely command the robot.

Then two different options can be considered for implementing the Scrum framework:

1. to define the sprints on the basis of the decomposition of the architecture into subsystems (Figure 3). Each sprint will address the development of one subsystem considered independently from the others, a supplementary sprint will address the integration of subsystems.
2. to define sprints on the basis of the phases of the robot mission. This means analyzing which parts of subsystems are invoked at each phase of the mission and define sprints aiming at incrementally developing these specific parts. Figure 4 shows the decomposition of each subsystem of the robot into parts. S1 and S2 are subsystems of the robot structure and motion subsystems, V1, V2 and V3 of the vision subsystem, HMI1 and HMI3 of the human machine interface subsystem. On the figure, we can see that the option taken here is to develop S1, V1 and HMI1 at one sprint, S2 and V2 at another sprint, and V3 and HMI3 in another sprint.

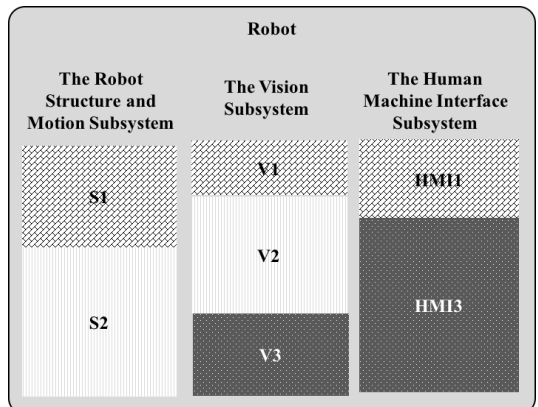

Figure 4: Distribution of the robot subsystems

These two options are presented here below.

### 4.1.1 Option 1

To define the sprints, the Product Owner has to prioritize the features inside the Product Backlog then plan each sprint with the Development Team. According to this project, we can distribute the Product Backlog using as a reference the global architecture defined in Figure 3, each sub-system is considered as a feature of the Product Backlog and shown in Figure 5.

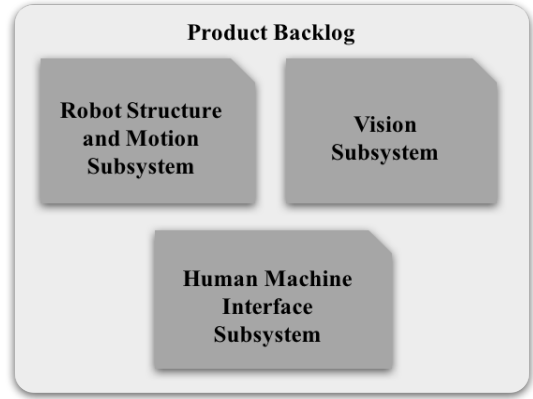

Figure 5: Product Backlog of the robot project

We thus propose four sprints, one sprint for each subsystem and a supplementary sprint for the assembly.

- Sprint 1: the first artifact is the architecture and motion system of the robot; it is the result of the first sprint.

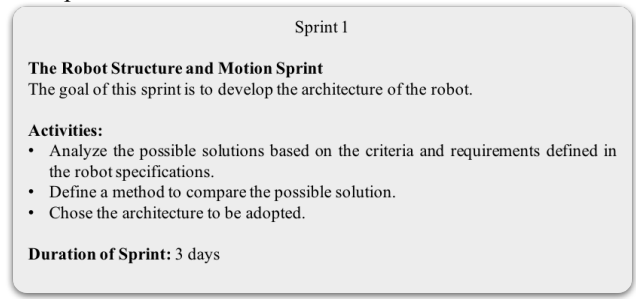

Figure 6: First sprint, the architecture and motion subsystem of the robot

- Sprint 2: the second artifact is the vision subsystem

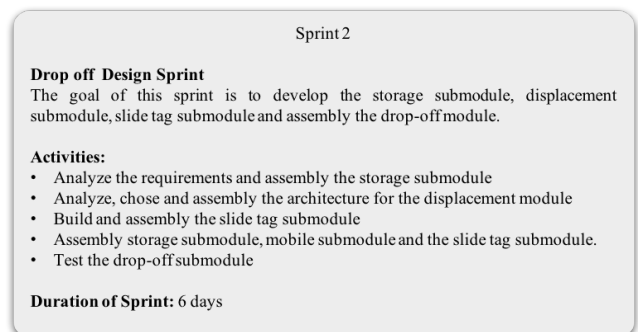

Figure 7: Second sprint, the vision subsystem

- <u>Sprint 3</u>: the human machine interface is the third increment of this sprint.

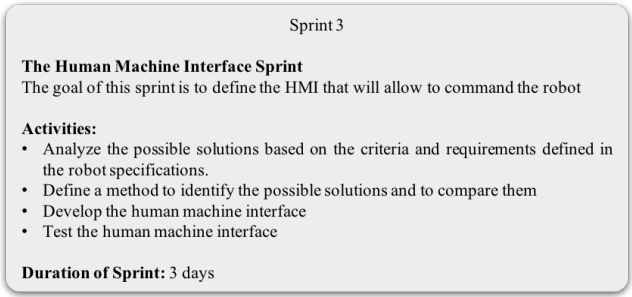

Figure 8: Third sprint, the human machine interface

- <u>Sprint 4</u>: this is the last sprint; the entire robot is the result of this sprint.

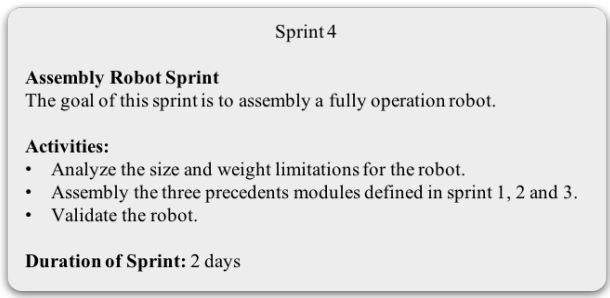

Figure 9: Fourth sprint, entire robot

The result of each sprint is considered as an increment, which means that once the first sprint is completed, the detail of the next phase (second sprint) is worked out, and so on. Each sprint has to follow a daily sprint meeting to know the status of the activities defined in each sprint, followed by the sprint review. If everything is OK, the development team can do the increment, otherwise the development team has to hold the sprint retrospective to correct or redefine the sprint. The proposal has to follow the framework described in Figure 2.

The focus point of this proposal is the architecture of the robot. The first three sprints follow a list of activities that allows to deliver and test a specific and independent part of the robot, each increment of these sprints are assembled in a fourth sprint to have a fully operational robot.

*4.1.2 Option 2*

This option proposes to identify which subsystem is invoked at each phase of the robot mission. To define the sprints, the Product Owner can consider the decomposition of the robot subsystems (Figure 4) as the Product Backlog then starts the sprint planning. Product Owner also has to prioritize the features inside the Product Backlog, then plans the sprints with the Development Team.

Once the sprints were defined, their development has to follow the framework described in Figure 2, in that case the increment is considered as the result of the integration of each subsystem to accomplish each phase of the global mission, sprints have to follow a daily sprint meeting and the sprint review as stablish in the scrum framework. To illustrate how sprints can be defined in this option, we presented the first sprint as follows:

- <u>Sprint 1</u>: the first artifact of this sprint is the integration of three subsystems to accomplish Phase 1. It needs the subsystem S1 (from the robot Structure motion and subsystem), V1 (from the Vision subsystem), and HMI1 (from the human machine interface subsystem).

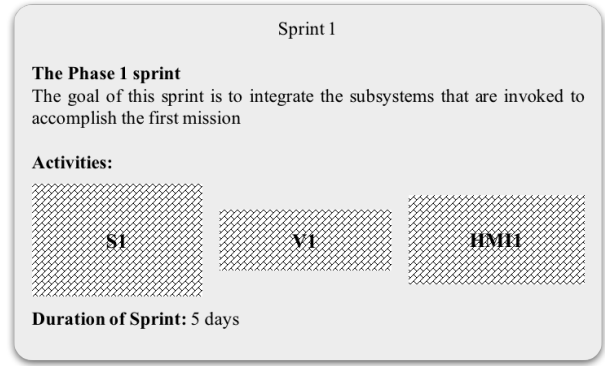

Figure 10: First sprint, the mobile module and HMI

Figure 10 shows the first sprint of this proposal. The activities were defined considering the subsystems needed and the integration of them is considered as the result of the first sprint (increment). The definition of the other two sprints should follow the same analysis.

This proposal stablishes a different way from the first one. After the roles definition, Product Owner identifies which subsystems were involved to complete each phase of the global mission then defines each sprint. It allows to deliver artifacts that contains hardware and software at the same time.

**4.2 Discussion**

Section 4.1 shows two different conceptual ways to implement the Scrum Framework as an alternative to the traditional management methods, the goal of this implementation is to develop a mobile connected robot in an agile way. The implementation was proposed following three phases that the robot has to follow to accomplish a global mission. In fact, this work is focused in order to evaluate if the development of the project robot is a good case study to implement the scrum framework, considering that this project not only includes software development but also hardware integration.

From the implementation of agile management on this educational project, we can find that developing the robot with Scrum could help students to deliver functional modules in each increment and reduce time for testing as presented. For example, they delivered functional parts of the robot including hardware and software at the same time in each sprint. The results of this proposal will be evaluated at the end of the school year.

This work also opens an additional discussion point relative to this investigation. The architecture and requirements were well defined before, but what if during the development that changes? As we know, requirements can be changed during the development; so, these changes may be discussed in the sprint review or in the sprint retrospective, but we do not really know if that will have a strong impact in the architecture. This point shall be evaluated next year, when agility will be implemented in new student projects.

Some difficulties could exist when trying to implement agile management methods on the mechanical development projects. For example, during the project, the project individual had more and more stress because of the self-management and self-organization rather than having direct orders from their superiors. From the management's view this creates uncertainty of project. The entire team is actually held accountable for the success or failure of a project even when the product owner is to be held responsible. It is not easy to divide the project into sprints according to the functions of product, this issue will be analyzed when students use the proposals stablished next year. Differences between the traditional management of the project and the implementation of an agile approach is another criterion to be considered as part of the analysis of this integration.

## 5 CONCLUSION

This article aimed at deploying, testing and evaluating the issues and benefits of an agile approach on a student project, in order to determine if this project could be a candidate to train students to a basic practice of agile management.

To this purpose, this paper considered and compared two options to implement agility. In the first option, four sprints were defined with a specific duration and related activities; at the end of each sprint, students can deliver a functional part of the robot and test it. In the second option, the sprint definition relied on the subsystems involved to achieve each phase of the global mission of the robot; it allows delivering software and hardware modules in parallel. During the sprint development, in both cases, close communication should be effective in order to cover all the requirements defined in the robot specifications. Specific roles should be considered to manage the evolution of the robot development; these roles allow the distribution of responsibilities during the development.

The paper concludes that implementing an agile method such as Scrum in this project during the system development could be interesting for students to improve cooperation among students, and quickly and efficiently develop the connected mobile robot, thus improving project performance. It also concludes that implementing Scrum in this project could allow to train students to basics of agile management.